\documentstyle[12pt,epsf,a4]{article}
\newcommand{\gsim}{\;\rlap{\lower 3.5 pt \hbox{$\mathchar \sim$}} \raise 1pt
 \hbox {$>$}\;}
\newcommand{\lsim}{\;\rlap{\lower 3.5 pt \hbox{$\mathchar \sim$}} \raise 1pt
 \hbox {$<$}\;}

\renewcommand{\thefootnote}{\fnsymbol{footnote}}
\begin{document}    

\title{\vskip-3cm{\baselineskip14pt
\centerline{\normalsize\hfill MPI/PhT/98--12}
\centerline{\normalsize\hfill TTP98--05}
%%%\centerline{\normalsize\hfill hep-ph/9802241}
\centerline{\normalsize\hfill February 1998}
}
\vskip.7cm
A theory driven analysis of the effective QED \\ coupling at $M_Z$
\vskip.5cm
}
\author{
 J.H.~K\"uhn$^{a}$
 and 
 M.~Steinhauser$^{b}$
}
\date{}
\maketitle

\begin{center}
$^a${\it Institut f\"ur Theoretische Teilchenphysik,
    Universit\"at Karlsruhe,\\ D-76128 Karlsruhe, Germany\\ }
  \vspace{3mm}
$^b${\it Max-Planck-Institut f\"ur Physik,
    Werner-Heisenberg-Institut,\\ D-80805 Munich, Germany\\ }
\end{center}
\vspace{1cm}

\begin{abstract}
  \noindent 
An evaluation of the effective QED coupling at the scale $M_Z$ is
presented.  It employs the predictions of perturbative QCD for the 
cross section of electron positron annihilation into hadrons up to order
$\alpha_s^2$, including the full quark mass dependence, and of order
$\alpha_s^3$ in the high energy region. This allows to predict the input
for the dispersion relations over a large part of the integration
region. The perturbative piece is combined with data for
the lower energies and the heavy quark thresholds. The result for the
hadronic contribution to the running of the coupling
$\Delta\alpha^{(5)}_{\rm had}(M_Z^2)= (277.5 \pm 1.7)\times 10^{-4}$
leads to
$(\alpha(M_Z^2))^{-1}=128.927 \pm 0.023$.
Compared to previous analyses the uncertainty is thus
significantly reduced.

%%%\noindent
%%%PACS numbers:
\end{abstract}

\vspace{.6cm}

%\thispagestyle{empty}
%\newpage
%\setcounter{page}{1}

%%%%%%%%%%%%%%%%%%%%%%%%%%%%%%%%%%%%%%%%%%%%%%%%%%%%%%%%%%%%
%%%%%%%%%%%%%%%%%%%%%%%%%%%%%%%%%%%%%%%%%%%%%%%%%%%%%%%%%%%%

\renewcommand{\thefootnote}{\arabic{footnote}}
\setcounter{footnote}{0}

The evolution of the electromagnetic coupling from its definition at
vanishing momentum transfer to its value at high energies constitutes
the dominant part of radiative corrections for electroweak
observables. The accurate determination of $\alpha(M_Z^2)$ is
thus essential for any precise test of the theory. At the same time the
indirect determination of the masses of heavy, hitherto unobserved
particles, e.g.~the Higgs boson or SUSY particles, depends critically on
this parameter. Of particular importance in this context is the hadronic
vacuum polarization. It is nearly as large as the leptonic contribution
but can only be related through dispersion relations to the cross
section for hadron production in electron positron annihilation, or more
conveniently to the familiar $R$ ratio. The integrand can thus be
obtained from data, phenomenological models and/or perturbative QCD (pQCD),
whenever applicable. A detailed analysis based on data and employing
pQCD above $40$~GeV has been performed in~\cite{EidJeg95} and their result
has been confirmed by subsequent 
studies~\cite{BurPie95,Swa95,AleDavHoe97}
following very similar strategies. Alternatively one may employ
pQCD also for lower energies, eventually as low as $2$~GeV as long as one
stays away from the quark thresholds. A first step in this direction has
been made in~\cite{MarZep95}. There, however, only the massless
approximation was employed for the normalization of the data.
Recently the ${\cal O}(\alpha_s^2)$ corrections
for $R$, including the full quark mass dependence, became 
available~\cite{HoaKueTeu95,CheKueSte9697,CheHarKueSte97}
which allows to extend pQCD
down relatively close to the respective thresholds for charm and bottom
production. These results have been used in~\cite{CheHoaKueSteTeu97}
to evaluate the perturbative contribution to the vacuum 
polarization. In this work we
complete the evaluation by incorporating those contributions which
cannot be obtained from pQCD: from the low energy region below roughly
$2$~GeV, the charmonium and bottomonium resonances and from the charm
threshold. The (re)normalization of the data from the 
PLUTO~\cite{PLUTO},
DASP~\cite{DASP}, 
and MARK~I collaborations~\cite{MARK1} will be based on the requirement
that they agree with pQCD for $\sqrt{s}\leq 3.7$~GeV and 
$\sqrt{s}\geq 5$~GeV. 
A similar analysis has been recently performed 
in~\cite{DavHoe97}
which is also based on pQCD, in particular the results
of~\cite{CheKueSte9697}
and which, furthermore, provides additional justification for the
applicability of pQCD at very low energies around $2$~GeV.
We will comment on the differences between this analysis
and~\cite{DavHoe97}, whenever appropriate, below.

Let us briefly describe the theoretical input for
our evaluation. The hadronic contribution
$\Delta \alpha^{(5)}_{\rm had}$ to the running from the
static limit to $M_Z$ is given by
\begin{eqnarray}
\Delta\alpha_{\rm had}^{(5)}(M_Z^2)
&=&
-\frac{\alpha M_Z^2}{3\pi}\,\mbox{Re}\,
\int_{4m_\pi^2}^\infty\,{\rm d}
s\,\frac{R(s)}{s\left(s-M_Z^2-i\epsilon\right)}
\,,
\label{eqdelal}
\end{eqnarray}
where the superscript ``$(5)$'' indicates that the top quark
is not included in the integral.
For the analysis we use $\alpha=\alpha(0)=1/137.0359895$ and 
$M_Z=91.187$~GeV.
It leads, after resummation of the leading
logarithms, to the following shift
\begin{eqnarray}
\alpha(s) &=& \frac{\alpha(0)}
      {1-\Delta\alpha_{\rm lep}(s)
        -\Delta\alpha^{(5)}_{\rm had}(s)
        -\Delta\alpha_{\rm top}(s)}
\,.
\end{eqnarray}
The quantity $R(s)$ can be experimentally
determined through a measurement of the total
cross section for electron positron annihilation
into hadrons. From the theoretical side it is
defined through the absorptive part of the
electromagnetic current correlator
\begin{eqnarray}
\left(-q^2g_{\mu\nu}+q_\mu q_\nu\right)\,\Pi(q^2)
&=&
i\int dx\,e^{iqx}\langle 0|Tj_\mu(x) j_\nu(0)|0 \rangle \,,
\label{eqpivadef}
\\
R(s)&=&12\pi\,\mbox{Im}\,\Pi(q^2=s+i\epsilon)
\,.
\end{eqnarray}
It can be calculated in the framework of
pQCD up to order $\alpha_s^3$
if quark masses are neglected~\cite{GorKatLar91SurSam91} and up to
${\cal O}(\alpha_s^2)$ with full quark mass 
dependence~\cite{HoaKueTeu95,CheKueSte9697,CheHoaKueSteTeu97}.
In this work pQCD will be assumed to be valid
above $1.8$~GeV (alternatively $2.125$~GeV).
In view of the validity of pQCD in tau lepton decays not only for the
total rate but also for the spectral function toward the upper 
end~\cite{Hoe97Wer} the substitution of inprecise data by pQCD seems
justified. 
Additional support for this approach can also be drawn from the analysis of
data for $R(s)$ below $1.8$~GeV~\cite{DavHoe97}. pQCD has also been
used in~\cite{AdeInd95} in the present context even down to
$1.4$~GeV.
The specific choice of $1.8$~GeV (or $2.125$~GeV) for the matching
between data and theory is dictated by the available data 
analysis~\cite{AleDavHoe97,DavHoe97} which we adopt for the present purpose.

Also excluded from this theory driven treatment are
the threshold region for charmed mesons (the interval from
$3.7$~GeV to $5$~GeV) and, similarly, for bottom mesons 
($10.5$~GeV to $11.2$~GeV), 
and the narrow charmonium and bottomonium resonances, where we use
the currently available data.
Perturbative QCD is even applicable in the charm and bottom threshold
regions, as far as the light quark contributions are concerned.
In the bottom threshold region we will therefore use data for
the bottom contribution only. 
%%%Therefore data will only be used to evaluate the
%%%charm and bottom pieces, respectively.

In the perturbative regions one receives contributions from light ($u$,
$d$ and $s$) quarks whose masses are neglected throughout, and from
massive quarks which demand a more refined treatment. Below the charm
threshold the light quark contributions are evaluated in order
$\alpha_s^3$ plus terms of order
$\alpha_s^2\, s/(4M^2_c)$
from virtual massive quark loops. Above $5$~GeV the full $M_c$ dependence is
taken into account up to order $\alpha^2_s$, and in addition the
dominant cubic terms in the strong coupling are incorporated, as well as
the corrections from virtual bottom quark loops of order
$\alpha_s^2\, s/(4M^2_b)$. Above $11.2$~GeV the same formalism
is applied to the massive bottom quarks and charmed mass effects are
taken into account through their leading contributions in an $M^2_c/s$
expansion.
All formulae are available for arbitrary renormalization scale $\mu$
which allows to test the scale dependence of the final answer. This
will be used to estimate the theoretical uncertainties
from uncalculated higher orders. Matching of $\alpha_s$
between the treatment with
$n_f=3$, $4$ and $5$ flavours is performed at the respective threshold
values. 
The influence of
the small ${\cal O}(\alpha^3_s)$ singlet piece which prevents a
clear separation of contributions from different quark species can safely be
ignored for the present purpose.
The details of the formalism can be found 
in~\cite{CheHoaKueSteTeu97}.

In Tab.~\ref{tabpert}, adopted 
from~\cite{CheHoaKueSteTeu97}, the perturbative hadronic contributions
are listed separately for a variety of energy intervals.
As our default values we adopt $\mu^2 = s$, $\alpha^{(5)}_s(M_Z^2)=0.118$,
$M_c=1.6$~GeV and $M_b=4.7$~GeV. In separate columns we list the
variations with a change in the renormalization scale, the strong
coupling constant and the quark masses:
\begin{eqnarray}
\delta\alpha_s\,\,=\,\,\pm0.003,\quad
\delta M_c\,\,=\,\,\pm0.2~{\mbox GeV},\quad
\delta M_b\,\,=\,\,\pm0.3~{\mbox GeV}.\quad
\label{eqdelta}
\end{eqnarray}

\begin{table}[t]
\renewcommand{\arraystretch}{1.3}
\begin{center}
{\small
\begin{tabular}{|l|r|r|r|r|r|}
\hline\hline
Energy range (GeV) & central value &$\delta\mu$ &
$\delta\alpha_s$ & $\delta M_c$ & $\delta M_b$\\ \hline
$  1.800-  2.125$&$    5.67$&$    0.22$&$    0.04$&$    0.00$&$    0.00$\\
$  2.125-  3.000$&$   11.66$&$    0.21$&$    0.06$&$    0.01$&$    0.00$\\
$  3.000-  3.700$&$    7.03$&$    0.06$&$    0.03$&$    0.00$&$    0.00$\\
$  1.800-  3.700$&$   24.36$&$    0.48$&$    0.13$&$    0.01$&$    0.01$\\
\hline
$  5.000-  5.500$&$    5.44$&$    0.03$&$    0.03$&$    0.06$&$    0.00$\\
$  5.500-  6.000$&$    4.93$&$    0.03$&$    0.02$&$    0.04$&$    0.00$\\
$  6.000-  9.460$&$   25.45$&$    0.11$&$    0.08$&$    0.10$&$    0.00$\\
$  9.460- 10.520$&$    5.90$&$    0.02$&$    0.01$&$    0.01$&$    0.00$\\
$ 10.520- 11.200$&$    3.48$&$    0.01$&$    0.01$&$    0.00$&$    0.00$\\
$  5.000- 11.200$&$   45.20$&$    0.19$&$    0.15$&$    0.21$&$    0.01$\\
(without $b\bar{b}$) &&&&&\\
\hline
$ 11.200- 11.500$&$    1.63$&$    0.00$&$    0.01$&$    0.00$&$    0.00$\\
$ 11.500- 12.000$&$    2.62$&$    0.00$&$    0.01$&$    0.00$&$    0.00$\\
$ 12.000- 13.000$&$    4.93$&$    0.01$&$    0.01$&$    0.00$&$    0.00$\\
$ 13.000- 40.000$&$   72.92$&$    0.08$&$    0.12$&$    0.02$&$    0.02$\\
$ 12.000- 40.000$&$   77.85$&$    0.09$&$    0.14$&$    0.02$&$    0.02$\\
$ 40.000-\infty$&$   42.67$&$    0.03$&$    0.06$&$    0.00$&$    0.00$\\
$ 11.200-\infty$&$  124.77$&$    0.12$&$    0.21$&$    0.03$&$    0.02$\\
\hline
$1.8-\infty$ (pQCD) & $  194.33$&$    0.79$&$    0.49$&$    0.24$&$    0.03$\\
\hline
QED & $    0.11$ & -- & -- & -- & -- \\
\hline\hline
%%%(without $b\bar{b}$) &&&&&\\
\end{tabular}
}
\parbox{14.cm}{\small
\caption{\label{tabpert}
Contributions to $\Delta\alpha^{(5)}_{\rm had}(M_Z^2)$ 
(in units of $10^{-4}$) from the energy
regions where pQCD is used (adopted from~\protect\cite{CheHoaKueSteTeu97}).
For the QED corrections the same intervals
have been chosen. For the variation of $\alpha_s(M_Z^2)$, $M_c$ and $M_b$
Eqs.~(\ref{eqdelta}) have been used. $\mu$ has been varied between 
$\protect\sqrt{s}/2$ and $2\protect\sqrt{s}$.
}}
\end{center}
\end{table}

In principle the theoretical tools are available to include
in the perturbative analysis the QED
corrections of order $\alpha$ and even $\alpha\alpha_s$. 
The relative size of the dominant correction
is estimated as
\begin{eqnarray}
\frac{\sum_i Q_i^4}{\sum_i Q_i^2} \frac{3}{4}\frac{\alpha}{\pi}
&\approx& \left(0.6 - 0.7\right) \times 10^{-3}
\,, 
\end{eqnarray}
and is also included in Tab.~\ref{tabpert}.

Perturbative QCD is clearly inapplicable in the charm threshold region
between $3.7$ and $5$~GeV where rapid variations of the cross section are
observed. Data have been taken more than $15$ years ago by the 
PLUTO~\cite{PLUTO},
DASP~\cite{DASP}, and MARK~I collaborations~\cite{MARK1}. 
The systematic errors of $10$ to
$20$~\% exceed the statistical errors significantly and are reflected in a
sizeable spread of the experimental results. To arrive at a reliable
evaluation of the charm contribution from this region, we adjust
the normalization
of the data (for each experiment individually)
to the theoretical predictions at the upper and lower endpoints
as follows: Data below and up to $3.7$~GeV are combined to determine the
factor $n_-$ which characterizes the mismatch between theory and
experiment below threshold
\begin{eqnarray}
n_- &\equiv& \left< \frac{R_{exp}(s)}{R_{pQCD}(s)} \right>
\,,
\end{eqnarray}
and an averaged experimental $R$ value just below threshold
\begin{eqnarray}
R_- &\equiv& n_-\, 
%%%\left< 
R_{pQCD}((3.7 {\rm GeV})^2) 
%%%\right>
\,.
\end{eqnarray}
In a similar way $n_+$ and $R_+$ are derived from the data around and above
$5$~GeV. The normalization factors  and the combined $R$ values as given
in Tab.~\ref{tabexp} are consistent with the systematic errors quoted by the
experiments. To account for the difference between $n_-$ and $n_+$ even
within one experiment, two models are used for the interpolation into the
interior of the interval. {\it Model~1:} The difference is due to the
different efficiencies for final states with and without charmed mesons.
{\it Model~2:} The difference is due to a linear $s$
dependence of the
experimental normalization and thus reflected in a linear $s$ dependence
of the quantity $1/n(s)$ in the threshold region.

\begin{table}[t]
\renewcommand{\arraystretch}{1.3}
\begin{center}
{\small
\begin{tabular}{|l|r|r|r|}
\hline\hline
 & PLUTO & DASP & MARK1\\ \hline 
\hline
Interval below (GeV) & $  3.6000 -   3.6600$ & $  3.6025 -   3.6500$ & $
  3.0000 -   3.6500$ \\ 
$n_-$ & $    1.04 \pm     0.01$ & $    1.06 \pm     0.02$ & $     1.18
 \pm     0.04$\\ 
$R_-$ & $    2.25 \pm     0.03$ & $    2.29 \pm     0.04$ & $     2.55
 \pm     0.08$\\ 
\hline
Interval above (GeV) & $  4.9800 -   4.9800$ & $  5.0000 -   5.1950$ & $
  5.1000 -   6.0000$ \\ 
$n_+$ & $    1.04 \pm     0.01$ & $    1.15 \pm     0.01$ & $
    1.12 \pm     0.01$\\ 
$R_+$ & $    3.85 \pm     0.04$ & $    4.27 \pm     0.04$ & $
    4.14 \pm     0.05$\\ 
\hline
$\Delta\alpha^{(5)}_{c\bar{c}}(M_Z^2)\times 10^{4}$ (Model 1) &
$   15.65 \pm     0.19$ & $   15.26 \pm     0.25$ & $
   16.22 \pm     0.32$\\ 
$\Delta\alpha^{(5)}_{c\bar{c}}(M_Z^2)\times 10^{4}$ (Model 2) &
$   15.64 \pm     0.16$ & $   15.68 \pm     0.30$ & $
   15.83 \pm     0.36$\\ 
\hline\hline
\end{tabular}
}
\parbox{14.cm}{\small
\caption{\label{tabexp}
Contribution to $\Delta\alpha^{(5)}_{\rm had}(M_Z^2)$
from the energy interval $3.7$ to $5.0$~GeV.
}}
\end{center}
\end{table}

The average of the two slightly different results is taken as
central value and the three experiments are then assumed to be uncorrelated.
The typical spread of $\pm0.2$ is taken as systematical uncertainty
which is combined linearly with the statistical error
for which we take the maximum of {\it Model~1} and {\it Model~2}.
The combined result is thus given by
\begin{eqnarray}
\Delta\alpha^{(5)}_{c\bar{c}}(M_Z^2) &=& 
\left(15.67 \pm 0.34\right)\times 10^{-4}
\,.
\end{eqnarray}
A similar approach has been adopted in~\cite{MarZep95}. 
In~\cite{MarZep95}, however, only MARK~I data were employed
(with DASP, PLUTO and Crystal Ball data used for cross checks), 
an energy independent correction factor was chosen, and the pQCD 
prediction for massless quarks was used for the calibration below charm
threshold.

For the three lowest and narrow charmonium resonances we use the narrow
width approximation:
\begin{eqnarray}
\Delta\alpha^{(5)}_R(M_Z^2) &=& 
                   \frac{3}{\alpha}\left(\frac{\alpha}{\alpha(M_R^2)}\right)^2
                   \frac{M_Z^2}{M_R^2}
                   \frac{M_R\Gamma_{ee}}{M_Z^2-M_R^2}
\,,
\label{delalres}
\end{eqnarray}
with $(\alpha/\alpha(M^2_\psi))^2=0.96$.
$M_R$ is the mass of the resonance and
$\Gamma_{ee}$ the partial width into electrons.
The errors from the
three charmonium resonances are added linearly.
The result, given in Tab.~\ref{tabfull}, differs by about 
$0.7\times 10^{-4}$ from~\cite{AleDavHoe97,DavHoe97}.

The contributions from the three lowest $\Upsilon$ resonances are
evaluated through Eq.~(\ref{delalres}) with 
$(\alpha/\alpha(M^2_\Upsilon))^2= 0.93$.
For the bottom threshold an approach similar to the one for charm could
be employed. However, the $b\bar{b}$ 
cross section between $10.5$ and $11.075$~GeV is
saturated by the three $\Upsilon$ resonances at $10.580$~GeV, $10.865$~GeV
and $11.019$~GeV. (The result for the six $\Upsilon$ resonances as listed
in Tab.~\ref{tabfull} differs from~\cite{DavHoe97} by $0.2\times10^{-4}$.)
For energies above $11.2$~GeV the perturbative
treatment seems adequate. Between $11.075$ and $11.2$~GeV a linear increase
from zero to the perturbative value is assumed, and the error is
conservatively taken to be equal to this value.

For the low energy region up to $1.8$~GeV we use the value
$(56.90\pm1.10)\times 10^{-4}$~\cite{DavHoe97}.
The individual results for the different regions are listed in 
Tab.~\ref{tabfull}.
Combining the experimental errors, those from $\alpha_s^{(5)}(M_Z^2)$, 
the quark masses and the theoretical error in quadrature, we find
\begin{eqnarray}
\Delta\alpha^{(5)}_{\rm had}(M_Z^2) &=& 
\left(277.45 \pm 1.68\right)\times 10^{-4}
\end{eqnarray}
as our main result.
Alternatively we could have used the more conservative analysis 
of~\cite{AleDavHoe97} for the region up to $2.125$~GeV with a 
contribution of $(63.42\pm2.59)\times 10^{-4}$ and pQCD only 
above $2.125$~GeV. The result of this approach, 
$\Delta\alpha^{(5)}_{\rm had}(M_Z^2) = (278.30\pm2.82)\times 10^{-4}$,
would differ slightly in the central value and significantly in the size
of the error.

\begin{table}[t]
\renewcommand{\arraystretch}{1.3}
\begin{center}
{\small
\begin{tabular}{|l|l|r|}
\hline\hline
Input & energy region & $\Delta\alpha^{(5)}\times 10^4$ \\ 
\hline
low energy data~\cite{DavHoe97} & $2m_\pi-1.8$~GeV & $   56.90 \pm     1.10$\\
narrow charmonium resonances & $J/\Psi, \Psi(2S), \Psi(3770)$ &
 $    9.24 \pm     0.74$\\
``normalized'' data & $3.7-    5.0$~GeV & $   15.67 \pm     0.34$\\
$\Upsilon$ resonances & $\Upsilon(1S)-\Upsilon(11.019)$ &
 $    1.17 \pm     0.09$\\
interpolation of $b\bar{b}$ & $11.075-11.2$ & $    0.03 \pm     0.03$\\
pQCD (and QED) & $1.8-\infty $ & $  194.45 \pm 0.96$\\
\hline
& total & $  277.45 \pm 1.68$\\ 
\hline\hline
\end{tabular}
}
\parbox{14.cm}{\small
\caption{\label{tabfull}
Contributions to $\Delta\alpha^{(5)}_{\rm had}(M_Z^2)$
from different energy regions.
}}
\end{center}
\end{table}

Frequently the contribution from the top quark is added to this value.
Using the three-loop QCD corrections~\cite{CheKueSte9697}
\begin{eqnarray}
\Delta\alpha_{\rm top} (s) \!&\!=\!&\!
-\frac{4}{45}\frac{\alpha}{\pi}\frac{s}{M_t^2}
\left\{ 1 + 5.062 \frac{\alpha_s^{(5)}(\mu^2)}{\pi}
         + \left(28.220 + 9.702\ln\frac{\mu^2}{M_t^2}\right)
           \left(\frac{\alpha_s^{(5)}(\mu^2)}{\pi}\right)^2
\label{eqtop}
\right.\\&&\mbox{}\left.
\!+\frac{s}{M_t^2}\left[
0.1071 + 0.8315 \frac{\alpha_s^{(5)}(\mu^2)}{\pi}
         + \left(6.924 + 1.594\ln\frac{\mu^2}{M_t^2}\right)
           \left(\frac{\alpha_s^{(5)}(\mu^2)}{\pi}\right)^2
\right]
\right\}
\,,
\nonumber
\end{eqnarray}
one obtains
\begin{eqnarray}
\Delta\alpha_{\rm top}(M_Z^2) &=& 
\left(-0.70 \pm 0.05 \right) \times 10^{-4}
\,, 
\end{eqnarray}
where we have used $M_t=175.6\pm5.5$~GeV.
For convenience Eq.~(\ref{eqtop}) is expressed in terms of 
$\alpha_s^{(5)}(\mu^2)$. For the numerical evaluation 
$\alpha_s^{(5)}(M_Z^2)=0.118$
has been chosen.

A major contribution to the vacuum polarization originates from the
leptons. The dominant term is given by
\begin{eqnarray}
\Delta\alpha_{\rm lep}(M_Z^2) &=&
\frac{\alpha}{\pi} \sum_{i\in\{e,\mu,\tau\}} 
\left(-\frac{5}{9} + \frac{1}{3}\ln\frac{M_Z^2}{m_i^2} - 2\frac{m_i^2}{M_Z^2} 
+{\cal O}\left(\frac{m_i^4}{M_Z^4}\right)
\right)
\nonumber\\&&\mbox{}
+\Delta\alpha_{\rm lep,2l}(M_Z^2)
+{\cal O}\left(\alpha^3\right)
\nonumber\\&\approx&
314.19 \times 10^{-4}
+\Delta\alpha_{\rm lep,2l}(M_Z^2)
+{\cal O}\left(\alpha^3\right)
\,.
\end{eqnarray}
The two-loop correction~\cite{KalSab55}
\begin{eqnarray}
\Delta\alpha_{\rm lep,2l}(M_Z^2) &=&
\left(\frac{\alpha}{\pi}\right)^2 \sum_{i\in\{e,\mu,\tau\}} \left(
-\frac{5}{24} + \zeta(3) + \frac{1}{4}\ln\frac{M_Z^2}{m_i^2}
+ 3\frac{m_i^2}{M_Z^2}\ln\frac{M_Z^2}{m_i^2}
+{\cal O}\left(\frac{m_i^4}{M_Z^4}\right)
\right)
\nonumber\\&\approx&
0.78 \times 10^{-4}
\end{eqnarray}
leads to a shift which could in principle become relevant in forthcoming
precision studies.

For the combined result we thus obtain
\begin{eqnarray}
\left(\alpha(M_Z^2)\right)^{-1} &=& 128.927 \pm 0.023,
\end{eqnarray}
if we use the more optimistic analysis~\cite{DavHoe97}
for the region below $1.8$~GeV, and alternatively
$\left(\alpha(M_Z^2)\right)^{-1} = 128.916 \pm 0.039$
if we employ the analysis from~\cite{AleDavHoe97} below $2.125$~GeV. 
In Tab.~\ref{tabcmp}
our result for $\Delta\alpha_{\rm had}^{(5)}(M_Z^2)$
is compared to earlier evaluations.
Our uncertainty is only a quarter of the one from the analysis 
of~\cite{EidJeg95} or \cite{AleDavHoe97} based on data 
only --- at the price of a more pronounced dependence on pQCD at 
relatively low energies. The
reduction of the error by a factor $1.5$ compared to~\cite{DavHoe97} 
is to a large extend a consequence of our different treatment of the charm
threshold.
The shift of the central value by $-1.0\times10^{-4}$
compared to~\cite{DavHoe97} is
mainly due to different
values for the charmonium and bottomonium contributions and 
our treatment of the charm threshold.
In the prediction for $\alpha(M_Z^2)$ this is partly
compensated by our inclusion of the leptonic two-loop contribution
of $0.8\times10^{-4}$.

\begin{table}[t]
\renewcommand{\arraystretch}{1.3}
\begin{center}
{\small
\begin{tabular}{|l|l|}
\hline\hline
$\Delta\alpha^{(5)}_{\rm had}(M_Z^2)\times 10^4$ & Reference \\
\hline
$273.2 \pm 4.2$   & \cite{MarZep95}, Martin et al. `95 \\
$280 \pm 7$       & \cite{EidJeg95}, Eidelman et al. `95\\
$280 \pm 7$       & \cite{BurPie95}, Burkhardt et al. `95\\
$275.2 \pm  4.6$  & \cite{Swa95},    Swartz `96\\
$281.7 \pm 6.2$   & \cite{AleDavHoe97}, Alemany et al. `97\\
$278.4 \pm 2.6^*$   & \cite{DavHoe97}, Davier et al. `97\\
$277.5 \pm 1.7$   & this work \\
\hline\hline
\end{tabular}
}
\parbox{14.cm}{\small
\caption{\label{tabcmp}
Comparison of different evaluations of $\Delta\alpha^{(5)}_{\rm had}(M_Z^2)$.
(${}^*\Delta\alpha_{\rm top}(M_Z^2)$ subtracted.)}}
\end{center}
\end{table}

Summary: The effective fine structure constant at $M_Z$ has
been evaluated with input from pQCD over most of the integration region.
A detailed analysis of the theoretical uncertainties has been
performed. The two-loop piece for leptons has been
included. In comparison with earlier results based on the analysis of
data a slight shift of the central value and a significant
reduction of the error has been obtained.

\vspace{2mm} 

\centerline{\bf Acknowledgments} 
\medskip\noindent
We would like to thank G.~Wolf for providing data of
the DASP Collaboration,
M.~Davier, A.~H\"ocker, T.~Teubner and D.~Zeppenfeld for helpful discussions,
and D. Karlen for pointing out an error in an earlier version of Tab.~2.
This work was supported by BMBF under Contract
057KA92P and DFG under Contract Ku \mbox{502/8-1}.


\begin{thebibliography}{99}

\bibitem{EidJeg95}
S. Eidelman and F. Jegerlehner, {\it Z. Phys.} {\bf C 67} (1995) 585.

\bibitem{BurPie95}
H. Burkhardt and B. Pietrzyk, {\it Phys. Lett.} {\bf B 356} (1995) 389.

\bibitem{Swa95}
M.L. Swartz, {\it Phys. Rev.} {\bf D 53} (1996) 5268.

\bibitem{AleDavHoe97}
R. Alemany, M. Davier and A. H\"ocker, 
{\it Eur. Phys. J.} {\bf C 2} (1998) 123.

\bibitem{MarZep95}
A.D. Martin and D. Zeppenfeld, {\it Phys. Lett.} {\bf B 345} (1995) 558.

\bibitem{HoaKueTeu95}
A.H. Hoang, J.H. K\"uhn and T. Teubner,
{\it Nucl. Phys.} {\bf B 453} (1995) 173.

\bibitem{CheKueSte9697}
K.G. Chetyrkin, J.H. K\"uhn and M. Steinhauser,  
{\it Phys. Lett.} {\bf B 371} (1996) 93;
{\it Nucl. Phys.} {\bf B 482} (1996) 213;
{\it Nucl. Phys.} {\bf B 505} (1997) 40.

\bibitem{CheHarKueSte97}
K.G. Chetyrkin, R. Harlander, J.H. K\"uhn and  M. Steinhauser,
{\it Nucl. Phys.} {\bf B 503} (1997) 339.

\bibitem{CheHoaKueSteTeu97}
K.G. Chetyrkin, A.H. Hoang, J.H. K\"uhn, M. Steinhauser and T. Teubner,
{\it Eur. Phys. J.} {\bf C 2} (1998) 137.

\bibitem{PLUTO}
J. Burmester et al. (PLUTO Coll.), {\it Phys. Lett.} {\bf B 66} (1977) 395;\\
L. Criegee and G. Knies, {\it Phys. Rep.} {\bf C 83} (1982) 151.

\bibitem{DASP}
R. Brandelik et al. (DASP Coll.), {\it Phys. Lett.} {\bf B 76} (1978) 361;\\
A. Petersen, Ph.D. thesis, Hamburg University (1978).

\bibitem{MARK1}
J.L. Siegrist et al. (MARK~I Coll.), {\it Phys. Rev.} {\bf D 26} (1982) 969.

\bibitem{DavHoe97}
M. Davier and A. H\"ocker, 
{\it Phys. Lett.} {\bf B 419} (1998) 419.

\bibitem{GorKatLar91SurSam91}
S.G. Gorishny, A.L. Kataev and S.A. Larin, 
{\it Phys. Lett.} {\bf B 259} (1991) 144;\\
L.R. Surguladze and M.A. Samuel,  
{\it Phys. Rev. Lett.} {\bf 66} (1991) 560;
(E) ibid., 2416;\\
K.G. Chetyrkin, {\it Phys. Lett.} {\bf B 391} (1997) 402.

\bibitem{Hoe97Wer}
A. H\"ocker (ALEPH Coll.), LAL-97-18;\\
S. Menke and N. Wermes (OPAL Coll.), private communication.

\bibitem{AdeInd95}
K. Adel and F.J. Yndur\'ain, Report Nos. FTUAM-95-32 and hep-ph/9509378. 

\bibitem{KalSab55}
G. K\"all\'en and A. Sabry, {\it K. Dan. Vidensk. Selsk. Mat.-Fys. Medd.} 
{\bf 29} (1955) No. 17.


\end{thebibliography}
\end{document}